\documentclass[aps,prl,twocolumn,superscriptaddress]{revtex4}
\usepackage{mathrsfs}
\usepackage{graphicx}
\usepackage{latexsym}
\usepackage{stmaryrd}
\usepackage{amssymb}
\usepackage{wasysym}
\usepackage{longtable}
\begin{document}

% Use the \preprint command to place your local institutional report
% number in the upper righthand corner of the title page in preprint mode.
% Multiple \preprint commands are allowed.
% Use the 'preprintnumbers' class option to override journal defaults
% to display numbers if necessary
%\preprint{}

%Title of paper
\title{Vortex-Antivortex annihilation dynamics in a square mesoscopic superconducting cylinder}

% repeat the \author .. \affiliation  etc. as needed
% \email, \thanks, \homepage, \altaffiliation all apply to the current
% author. Explanatory text should go in the []'s, actual e-mail
% address or url should go in the {}'s for \email and \homepage.
% Please use the appropriate macro foreach each type of information

% \affiliation command applies to all authors since the last
% \affiliation command. The \affiliation command should follow the
% other information
% \affiliation can be followed by \email, \homepage, \thanks as well.
%\author{Edson Sardella}
%\email{sardella@fc.unesp.br}
%\author{Andr\'e Luiz Malvezzi}
%\email{malvezzi@fc.unesp.br}
%\author{Paulo Noronha Lisboa-Filho}
%\email{plisboa@fc.unesp.br}
%\email[]{Your e-mail address}
%\homepage[]{Your web page}
%\thanks{}
%\altaffiliation{}
\author{Edson Sardella}
\author{Paulo Noronha Lisboa Filho}
\affiliation{ UNESP-Universidade Estadual Paulista, Departamento de
F\'{\i}sica, Faculdade de Ci\^encias, Caixa Postal 473, 17033-360,
Bauru-SP, Brazil}
\author{Cl\'ecio C.\ de Souza Silva}
\author{Leonardo Ribeiro Eul\'alio Cabral}
\affiliation{
Departamento de F\'{\i}sica, Universidade Federal de Pernambuco, 50670-901, Recife-PE, Brazil}
\author{Wilson Aires Ortiz}
\affiliation{ Departamento de F\'{\i}sica, Universidade Federal de
S\~ao Carlos, 13565-905, S\~ao Carlos-SP, Brazil}

%Collaboration name if desired (requires use of superscriptaddress
%option in \documentclass). \noaffiliation is required (may also be
%used with the \author command).
%\collaboration can be followed by \email, \homepage, \thanks as well.
%\collaboration{}
%\noaffiliation

\date{\today}

\begin{abstract}

The dynamics of the annihilation of a vortex-antivortex pair is
investigated. The pair is activated magnetically during the run of a
simulated hysteresis loop on a square mesoscopic superconducting
cylinder with an antidot inserted at its center. We study the
nucleation of vortices and antivortices by first increasing the
magnetic field, applied parallel to the axis of the sample, from
zero until the first vortex is created. A further increase of the
field pulls the vortex in, until it reaches the antidot. As the
polarity of the field is reversed, an antivortex enters the scene,
travels toward the center of the sample and eventually the pair is
annihilated. Depending on the sample size, its temperature, and
Ginzburg-Landau parameter, the vortex-antivortex encounter takes
place at the antidot or at the superconducting sea around it. The
position and velocity of the vortex and antivortex singularities
were evaluated as a function of time. The current density,
magnetization and order parameter topology were also calculated.
\end{abstract}

% insert suggested PACS numbers in braces on next line
% insert suggested keywords - APS authors don't need to do this
%\keywords{}

%\maketitle must follow title, authors, abstract, \pacs, and \keywords
\maketitle %\widetext

%\section{Introduction}
%Here goes the introduction.

%\section{The Theoretical Formalism}
%Here goes the theoretical formalism.

%\section{Results and Discussion}
%Here goes the results and the discussion.

Achieving a deep understanding of the nucleation and propagation of
vortices in real superconductors is a truly complex task, since
these entities interact with almost everything: first, with the
surface of the specimen, to surpass it; upon entrance, with other
vortices that might have already penetrated, and also with defects,
which might attract them and even act as pinning centers. Additional
difficulties to emulate the problem arise from the fact that
vortices generate heat while propagating, what can be harmful to the
robustness of the superconducting properties, if not catastrophic,
as is the case of vortex avalanches observed in some superconducting
films \cite{duran,leiderer2,johansen,colauto2,colauto,leiderer}. It
is quite common, however, that the existence of pinning potentials
represent a beneficial feature, since vortices can thus be prevented
from undergoing dissipative motion. An interesting approach to the
problem, which enables one to address most specificities without
excessive complexity, is to work in the small universe of mesoscopic
samples. In such an ambient, one can accommodate the essential
ingredients: relatively important surface-to-volume ratio, only a
few vortices on scene, and a number of defects - the so-called
antidots - usually arranged in a regular pattern. Furthermore, one
can study the interaction of an individual vortex-antivortex (V-AV)
pair and, eventually, witness their mutual annihilation.

Recently, there has been many studies about V-AV configurations in
mesoscopic superconductors (see for instance References
\cite{francois1,francois2,moshalkov,melnikov,chibotaru}). The
authors of these references have found that vortices and
antivortices may coexist in equilibrium in configurations which look
like a V-AV molecule. A somewhat common approach is to assume an
\emph{a priori} configuration and minimize the free energy in terms
of some relevant parameter for which the V-AV molecule is a stable
configuration. Here, we will focus in a rather different approach
concerning more with the dynamics of a V-AV encounter. The aim of
the present work is to elucidate the details involved in the process
of creation of pairs, following their time-evolution and ultimate
disappearance. We opted to do this making no use of \emph{a priori}
assumptions regarding symmetries, but simply varying the applied
magnetic field, allowing for the spontaneous nucleation of a vortex
and, cycling the field, of an antivortex, which form a pair of
interacting entities, whose subsequent time-evolution is monitored.

In this Letter we present the dynamics of the annihilation of a V-AV
pair activated magnetically during a short hysteresis loop. The
geometry we consider is a square mesoscopic superconducting cylinder
with an antidot placed at its center. Simulations are made in the
presence of an applied magnetic field parallel to the axis of the
cylinder. We study the nucleation of vortices and antivortices by
first increasing the magnetic field from zero until the first vortex
is created. The field was then decreased toward negative values.
During this process the vortex travels heading the center of the
sample, is trapped by the antidot and is then annihilated by an
antivortex as the polarity of the applied field is reversed. To
monitor the whole process, we evaluate the position and velocity of
the vortex and antivortex singularities as a function of time. We
also calculate the order parameter topology, the current density,
and the magnetization. In order to follow the dynamics of the
physical quantities we use the gauge-invariant time-dependent
Ginzburg-Landau (TDGL) equations. These equations describe the
time-evolution of the complex order parameter $\psi$ and the vector
potential ${\bf A}$, which is related to the local magnetic field
through the expression ${\bf h}=\mbox{\boldmath $\nabla$}\times{\bf
A}$. In the absence of transport currents, these equations read

\begin{eqnarray}
\eta\frac{\partial\psi}{\partial t} & = & -\left (-i\mbox{\boldmath
$\nabla$}-{\bf A} \right )^2\psi+(1-T)\psi(1-|\psi|^2)\;,\nonumber \\
\beta\frac{\partial{\bf A}}{\partial t}& = & {\bf
J}_s-\kappa^2\mbox{\boldmath $\nabla$}\times{\bf h}\;,\label{tdgleq}
\end{eqnarray}
where the supercurrent density is
\begin{equation}
{\bf J}_s=(1-T)\Re\left [ \psi\left ( -i\mbox{\boldmath
$\nabla$}-{\bf A} \right )\psi \right ]\;.
\end{equation}

\begin{figure}[h]
  \centering
  \includegraphics[width=8.6cm]{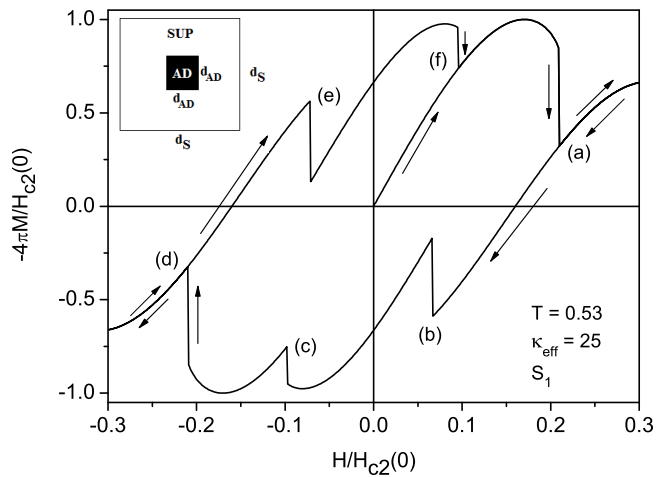}
  \caption{The magnetization curve as a function of the external applied magnetic
  field normalized to its maximum value. The meaning of the points marked in the
  figure are explained in the text. The arrows indicate the direction of the hysteresis loop.
  The inset illustrates the geometry of the problem.}\label{fig1}
\end{figure}

Here, the distances are measured in units of the coherence length at
zero temperature $\xi(0)$; the magnetic field in units of the upper
critical field at zero temperature $H_{c2}(0)$; the temperature in
units of the critical temperature $T_c$; the time is in units of the
characteristic time $t_0=\pi\hbar/8K_BT_c$ \cite{werthamer};
$\kappa$ is the Ginzburg-Landau parameter; $\eta$ and $\beta$ are
the relaxation times of the order parameter and the vector
potential, respectively. We have solved the TDGL equations upon
using the link variables method.\cite{gropp,buscaglia,sardella}
Since we consider invariance of the system along the $z$ direction,
our approach could only be applied to a square mesoscopic
superconducting cylinder. However, it might also be used for a very
thin superconducting film of thickness $d \le \xi(0)$ ($d \le 1$ in
dimensionless units), as long as we replace $\kappa^2$ in
Eq.~\ref{tdgleq} by $\kappa_{eff}=\kappa^2/d$.\cite{priour} The
geometry we have considered is depicted at the top left corner of
Fig.~\ref{fig1}: the lateral size of the sample is $d_S$, and of the
antidot $d_{AD}$.

For the study presented in this Letter, the relaxation times were
kept fixed at $\eta=\beta=1$. The GL parameter was also maintained
at $\kappa_{eff}=25$, corresponding, for instance, to a very thin
film of Nb with thickness $d\approx 7.3$ nm (assuming $\xi(0)=40$
nm, $\kappa=2.125$, $\kappa_{eff}=\kappa^2/(d/\xi(0))\approx 25$).

For $T=0.53$, Fig.~\ref{fig1} shows a short hysteresis loop which
was made for a mesoscopic superconducting square (hereafter referred
to as sample $S_1$) of dimensions $d_S=12$, with a small antidot at
the center, with size $d_{AD}=2$. Points marked with letters,
$(a),\ldots,(f)$ , indicate the values of $H$ where a vortex
(antivortex) either enters or exits the sample. The magnetic field
is increased from zero until a value somewhat above that at which
the first two vortices penetrate in $(a)$. It is then reversed until
the opposite value is achieved, and then reversed once more. At
point $(d)$, two antivortices penetrate the superconductor. Points
$(b)$ and $(e)$ correspond to the exit of a vortex and an
antivortex, respectively, immediately after which one flux quantum
still remains in the antidot. The point we will be mainly focusing
on is $(c)$ [or, equivalently, $(f)$]. This point corresponds to the
entrance of an antivortex, which will encounter a vortex trapped in
the antidot. The V-AV pair will annihilate either inside or outside
the antidot.

We have made a systematic study of the V-AV annihilation process,
varying the experimentally accessible relevant parameters: $T$, $d_S$, and $d_{AD}$.
The other parameters were kept fixed as specified above.

We turn now into a brief discussion of our results. Firstly, we
found that for $d_S$ and $d_{AD}$ fixed, the V-AV annihilation
occurs at a limited temperature interval. Above the upper limit of
such interval, the vortex trapped at the antidot leaves the sample
before the entrance of the antivortex, and no collision can possibly
occur. For the sample $S_1$ this upper threshold is somewhat above
$T = 0.53$. As a matter of fact, since the penetration length
increases as the temperature approaches $T_c$, the supercurrent
density associated with the pinned vortex spreads to distances large
enough to reach the external border of the sample. This enhances an
attractive interaction between the surface and the trapped vortex,
facilitating its escape.

\begin{figure}[h]
  \centering
  \includegraphics[width=8.6cm]{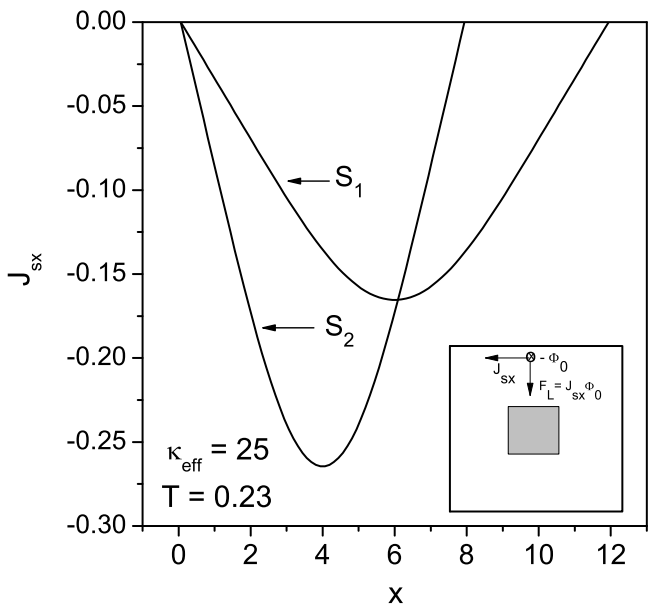}
  \caption{The supercurrent density along the top surface of the superconductor. The parameters used
  are specified in the figure. The inset shows the Lorentz force on the antivortex due to the vortex in the antidot.}\label{fig3}
\end{figure}

At a given temperature, there is a minimum width of the
superconducting frame to allow for a V-AV collision outside the
antidot. For example, for $T=0.23$ and $d_{AD}=2$ the minimum size
of the frame \cite{frame} is 6 (these dimensions define sample
$S_2$: $d_S=8$ and $d_{AD}=2$). Our reasoning is as follows: the
antivortex nucleated at the external border is kept pinned by the
surface barrier at the external border while, in turn, the vortex
remains pinned at the antidot. Both entities attract each other and
the interaction is larger for smaller superconducting frames. It is
then easier to pull the vortex out of the antidot for shorter
distances. On the other hand, if the superconductor width is
increased, the antivortex starts its excursion toward the center
before the V-AV attraction becomes appreciable. Being thus less
anchored by the surface, the antivortex travels all the way to the
antidot and the annihilation process completes there. In support to
this argument, Fig.~\ref{fig3} depicts the values of the $x$
component of the supercurrent density at the surfaces $y=12$ and
$y=8$ for both samples $S_1$ and $S_2$, respectively. These values
of $J_{sx}$ were calculated in the absence of applied field, $H=0$,
and thus represent the supercurrent due \emph{only} to the vortex
inside the antidot. The V-AV attraction (Lorentz force) will be
proportional to $J_{sx}$. Notice that, for both samples, the
absolute value of $J_{sx}$ is maximum at the center of the square
edge. One can clearly see that, in the middle of the sample edge,
$x=d_S/2$, the intensity of the Lorentz force is systematically
larger for the sample $S_2$ than for $S_1$.

\begin{figure}[h]
  \centering
  \includegraphics[width=8.6cm]{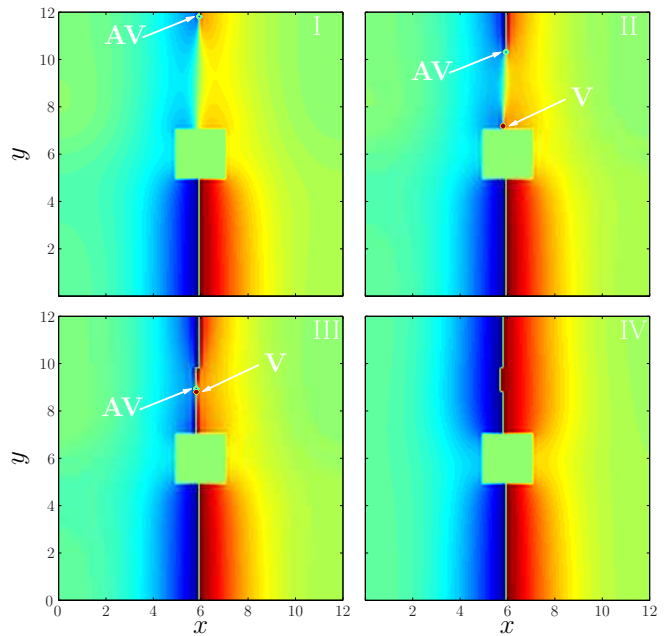}
  \caption{(Color online) The pictures show the topological shape of the phase of the order
  parameter; the phase goes from $-\pi$ (blue)  to $\pi$ (red). The last picture (IV) corresponds
  to the stationary state of point $(c)$ of the hysteresis loop of Fig.~\ref{fig1}.}\label{fig2}
\end{figure}

Fig.~\ref{fig2} shows the dynamics of the V-AV collision for the
sample $S_1$ at $T=0.53$. Immediately after entering the sample,
only the antivortex is in the superconducting sea. As it moves
forward to the antidot, at some instant the trapped vortex will come
out and both will collide and annihilate.

\begin{figure}[h]
  \centering
  \includegraphics[width=8.6cm]{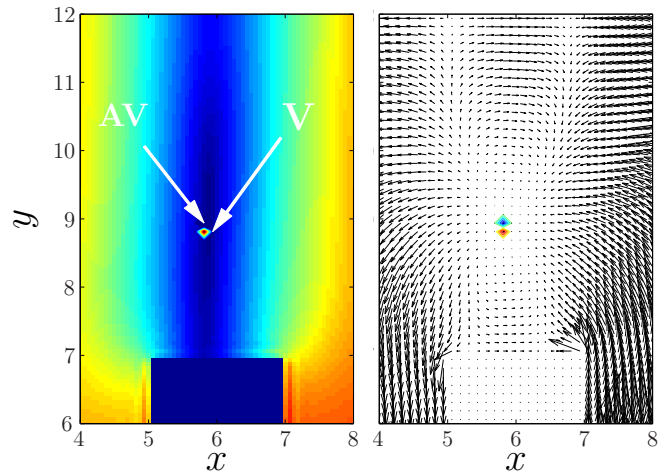}
  \caption{(Color online) The absolute vale of the order parameter (left) and the supercurrent
  distribution (right) for the sample $S_1$; very dark color means order parameter very depreciated.
  The size of the arrows (right) are not real; they were enlarged for better viewing.}\label{fig4}
\end{figure}

Another interesting characteristic we have found was that as the
antivortex penetrates, its interaction with the vortex is so strong
that both appear entirely deformed. Usually, vortices look like a
round flux filament surrounded by whirling supercurrents. However,
when the vortex and the antivortex approach one another while still
partially pinned at the corresponding surfaces, they become
alongated, as if squashed, and form a narrow channel between the
external surface and the antidot. This can be seen in
Fig.~\ref{fig4}, which represents the topology of the order
parameter $\psi$ and the supercurrent distribution at the very
moment when the collision takes place for the sample $S_1$ at
$T=0.53$. It is very suggestive that, during the collision time,
supercurrents nearly vanish accross the line of the elongated pair
V-AV, just like in a phase-slip event \cite{mooji}, which is the
dual process of Josephson tunneling.

\begin{figure}[h]
  \centering
  \includegraphics[width=8.6cm]{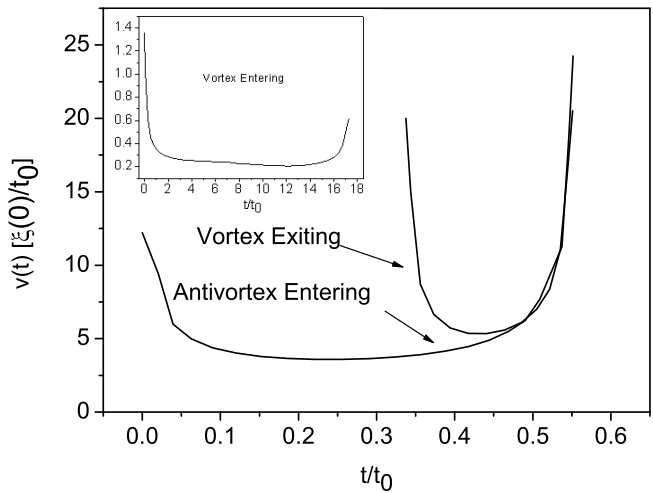}
  \caption{The modulus of the velocity of the vortex and the antivortex for
  the sample $S_1$ at $T=0.53$ as a function of time (point $(c)$ of Fig.~\ref{fig1}); the inset corresponds to point $(a)$.}\label{fig5}
\end{figure}

We also have estimated the instantaneous velocity of the vortex on
its way from nucleation at the surface through its capture by the
antidot (point $(a)$ of Fig.~\ref{fig1}). In addition, the
velocities of both the antivortex entering the sample and that of
the vortex exiting the antidot, on their route to  annihilation
(point $(c)$ of Fig.~\ref{fig1}), were also determined. The results
are shown in Fig.~\ref{fig5}. It is clearly seen that the velocities
achieved during the collision process are much larger than those of
a vortex penetrating the sample (at least an order of magnitude, as
shown in the inset).

In Reference \cite{sivakov} (see also \cite{andronov,doettinger}) it
was estimated that a phase-slip line can achieve a velocity which is
two orders of magnitude higher than an Abrikosov vortex which, in
turn, is one order of magnitude faster than Josephson vortices. As
Abrikosov vortices are quoted \cite{sivakov} to travel at $10^3$
m/s, this means that typical velocities for a phase-slip line are of
the order of  $10^5$ m/s. Although there is a conceptual difference
between the narrow channel of Fig.~\ref{fig4} (a composition of a
vortex and an antivortex) and a phase-slip line, the similarity in
format is reinforced by the fact that their velocities are
approximately the same. For instance, for the sample $S_1$, the
window during which the antivortex remains visible is $\Delta
t=0.5517t_0$ and the distance it travels is $\Delta y=2.875\xi(0)$;
inserting $T_c=3.72$ K and $\xi(0)=230$ nm (the relevant parameters
for Sn, which were used by Sivakov \emph{et al.} \cite{sivakov}),
the average velocity is $v_{\rm AV}=1.5\times 10^5$ m/s. On the
other hand, for the vortex exiting the antidot we obtain $\Delta
t=0.2135t_0$ and $\Delta y=1.625\xi(0)$; which gives $v_{\rm
Vs}=2.2\times 10^5$ m/s. Thus, in terms of format and velocity, the
annihilation process could be envisaged as a phase slip. It is worth
noticing that the large velocities anticipated for the V-AV pair
along the collision process, are similar to those developed during
the early stage of a vortex avalanche, as observed by the authors of
Ref.~\cite{leiderer}, who reported velocities as large as $1.8\times
10^5$ m/s for avalanches in YBCO films.

For the vortex first entering the antidot we found an average
velocity slightly larger then that for a normal Abrikosov vortex:
$\Delta t=17.2792t_0$, $\Delta y=4.125\xi(0)$, which yields $v_{\rm
Ve}=6.8\times 10^3$ m/s. A value somewhat larger - although in the
same order of magnitude - might be attributed to the additional
attraction exerted by the antidot.

In summary, we have studied the dynamics of the a V-AV annihilation.
The process was activated magnetically through a short hysteresis
magnetization loop. Our results indicate that the V-AV collision can
take place either inside the antidot or at the superconducting sea.
In addition, we have shown that, during the annihilation process,
vortices and antivortices may be stretched by a competition between
their mutual attraction and the forces that act to keep them
anchored at the surfaces. To complete the picture, we have also
reported velocities of the entities while travelling through the
superconducting frame of the sample.

\begin{acknowledgments}
The authors thank the Brazilian Agencies FAPESP, FACEPE, and CNPq
for financial support.
\end{acknowledgments}


\begin{thebibliography}{99}

\bibitem{duran} C.\ A.\ Duran, P.\ L.\ Gammel, R.\ E.\ Miller, and
D.\ J.\ Bishop, Phys.\ Rev. B {\bf 52}, 75 (1995).

\bibitem{leiderer2} P.\ Leiderer, J.\ Boneberg, P.\ Brull, V.\
Bujok, and S.\ Herminghaus, Phys. Rev. Lett.\ {\bf 71}, 2646 (1993).

\bibitem{johansen} T.\ H.\ Johansen, M.\ Baziljevich, D.\ V.\
Shantsev, P.\ E.\ Goa, Y.\ M.\ Galperin, H.\ N.\ Kang, H.\ J.\ Kim,
E.\ M.\ Choi, M.\ S.\ Kim, \emph{et al.}, Superconductor Science \&
Technology {\bf 14}, 726 (2001).

\bibitem{colauto2} F.\ Colauto, E.\ M.\ Choi, J.\ Y.\ Lee, S.\ I.\ Lee, V.\ V.\ Yurchenko, T.\
H.\ Johansen, and W.\ A.\ Ortiz, Superconductor Science \&
Technology {\bf 20}, L48 (2007).

\bibitem{colauto} F.\ Colauto, E.\ J.\ Patino, M.\ G.\ Blamire, and
W.\ O.\ Ortiz, Superconductor Science \& Technology {\bf 21}, 045018
(2008).

\bibitem{leiderer} B.\ Biehler, B.\-U.\ Runge, P.\ Leiderer, and R.\ G.\ Mints, Phys.\ Rev.\ B \textbf{72}, 024532 (2005); B.\-U.\ Runge, U.\ Bolz, J.\ Eisenmenger, and P.\ Leiderer, Physica C \textbf{341-348}, 2029 (2000).

\bibitem{francois1} G.\ R.\ Berdiyorov, M.\ V.\ Milo\v{s}evi\'c, and F.\ M.\ Peeters, Phys.\ Rev.\ Lett.\ \textbf{96}, 207001 (2006).

\bibitem{francois2} R.\ Geurts, M.\ V.\ Milo\v{s}evi\'c, and F.\ M.\, Peeters, Phys.\ Rev.\ Lett.\ \textbf{97}, 137002 (2006).

\bibitem{moshalkov} V.\ R.\ Misko, V.\ M.\ Fomin, J.\ T.\ Devreese, and V.\ V.\ Moshchalkov, Phys.\ Rev.\ Lett.\ \textbf{90}, 147003 (2003).

\bibitem{melnikov} A. S. Mel'nikov, I.\ M.\ Nefedov, D.\ A.\ Ryzhov, I.\ A.\ Shereshevskii, V.\ M.\ Vinokur, and P.\ P.\ Vysheslavtsev, Phys.\ Rev.\ B \textbf{65}, 140503(R) (2002).

\bibitem{chibotaru} L.\ F.\ Chibotaru, A.\ Ceulemans, V.\ Bruyndoncx, and V.\ V.\ Moshchalkov, Nature (London) \textbf{408}, 833 (2000).

\bibitem{werthamer} N.\ R.\ Weethamer in \emph{Superconductivity}, edited by R.\ D.\ Parks (New York, 1969).

\bibitem{gropp} W.\ D.\ Groop. H.\ G.\ Kaper, G.\ K.\ Leaf, D.\ M.\ Levine, M.\ Palumbo, and V.\ M.\ Vinokur, J.\ Comput.\ Phys.\ {\bf 123}, 254 (1996).

\bibitem{buscaglia} G.\ C.\ Buscaglia, C.\ Bolech, and A.\ L\'opez in \emph{Connectivity and Superconductivity}, edited by J.\ Berger e J.\ Rubinstein (Springer, Heidelberg, 2000).

\bibitem{sardella} E.\ Sardella, A.\ L.\ Malvezzi, P.\ N.\ Lisboa-Filho, and W.\ A.\ Ortiz, Phys.\ Rev.\ B \textbf{74},
014512 (2006).

\bibitem{priour} D.\ J.\ Priour and H.\ A.\ Fertig, Phys.\ Rev.\ B \textbf{67}, 054504 (2003).

\bibitem{frame} While changing sample and antidot sizes, we have
always adopted unitary steps.

\bibitem{mooji} J.\ E.\ Mooij, and Yu.\ V.\ Nazarov, Nature Physics \textbf{2}, 169 (2006).

\bibitem{sivakov} A.\ G.\ Sivakov, A.\ M.\ Glukhov, A.\ N.\ Omelyanchouk, Y.\ Koval, P.\ M\"uller, and A.\ V.\ Ustinov, Phys.\ Rev.\ Lett.\ {\bf 91}, 267001 (2003).

\bibitem{andronov} A.\ Andronov, I.\ Gordion, V.\ Kurin, I.\ Nefedov, and I. Shereshevsky, Physica C {\bf 213C}, 193 (1993).

\bibitem{doettinger} S.\ G.\ Doettinger, S.\ Kittelberger, R.\ P.\ Huebener, and C.\ C.\ Tsuei, Phys.\ Rev.\ B {\bf 56}, 14157 (1997).

\end{thebibliography}
\end{document}